\documentclass[runningheads]{llncs}
\usepackage{graphicx}
\usepackage[hidelinks]{hyperref}
\usepackage{times}
\usepackage{latexsym}
\usepackage{graphicx}
\usepackage{amsmath}
\usepackage{amsfonts}
\usepackage{amssymb}
\allowdisplaybreaks
\usepackage{booktabs}
\usepackage{tabularx}
\usepackage{subcaption}
\usepackage{caption}
\usepackage{multirow,bigdelim}

\begin{document}
\title{A Methodology for the Prediction of Drug Target Interaction using CDK Descriptors}
%
%
\author{Tanya Liyaqat\inst{1}
\and
Tanvir Ahmad\inst{1}
\and
Chandni Saxena\inst{2}} 
\authorrunning{T. Liyaqat et al.}
%
\institute{Jamia Millia Islamia University, New Delhi, India\\\email{tanyaliyaqat791@gmail.com} \\
\email{tahmad2@jmi.ac.in}\\ \and
The Chinese University of Hong Kong, Hong Kong SAR\\
\email{csaxena@cse.cuhk.edu.hk}}
\maketitle              
\begin{abstract}
Detecting probable Drug Target Interaction (DTI) is a critical task in drug discovery. Conventional DTI  studies are expensive, labor-intensive, and take a lot of time, hence there are significant reasons to construct useful computational techniques that may successfully anticipate possible DTIs. Although certain methods have been developed for this cause, numerous interactions are yet to be discovered, and prediction accuracy is still low. To meet these challenges, we propose a DTI prediction model built on molecular structure of drugs and sequence of target proteins. In the proposed model, we use Simplified Molecular-Input Line-Entry System (SMILES)  to create CDK descriptors, Molecular ACCess System (MACCS) fingerprints, Electrotopological state (Estate) fingerprints and amino-acid sequences of targets to get Pseudo Amino Acid Composition (PseAAC). We target to evaluate performance of DTI prediction models using CDK descriptors. For comparison, we use benchmark data and evaluate models' performance on two widely used fingerprints, MACCS fingerprints and Estate fingerprints. The evaluation of performances shows that CDK descriptors are superior at predicting DTIs. The proposed method also outperforms other previously published techniques significantly.

\end{abstract}
\section{Introduction}
Drug target interaction (DTI) is a prominent task in drug discovery and research. It entails detecting possible links among chemical compounds and protein targets which acts as a guide in the preliminary phases of drug discovery and developmental research. Experiments carried out in wet labs are labor intensive and require a significant amount of money~\cite{paul2010improve}. According to statistics, each novel molecular entity takes around 1.8 billion USD and the authorization of a novel drug application usually requires at least 9 years~\cite{dickson2004key}. As a result, high-efficiency computational prediction techniques to investigate drug target interactions based on Machine Learning (ML) and Deep Learning (DL) have sparked a lot of attention in recent years~\cite{chen2018machine}.
The bonding of a medicine to a target’s location resulting in the alteration of its functioning is considered as drug target interaction. Any chemical molecule that causes an alteration in the body's physiology when swallowed, ingested, or inhaled is referred to as a medication or medicine. On the other hand, targets consist of elements as nucleic acids or lipids, that are intended to modify. Ion channels, enzymes, nuclear receptors, and G-protein coupled receptors are among the most popular biological targets.
To treat illness and ailments, the medicine inhibits the target's function in order to prevent certain catalytic processes from occurring in the human body. This is accomplished by preventing it from interacting with particular enzymes known as substrates. The drug discovery procedure that detects novel therapeutic molecules for targets relies heavily on DTI prediction~\cite{sachdev2019comprehensive}.
Feature-based computational techniques for DTI prediction have gained significant attention over the years. The availability of the structural information of chemical compounds in the form of fingerprints or descriptors has played an important role. However, most studies consider fingerprints over descriptors. Hence, it becomes important to compare performance and identify better alternative. We provide more details about feature-based techniques in Section~\ref{sec2}.

Considering the widely accepted ability of structure information of molecules, we aim to evaluate the performance of CDK descriptors against two widely used fingerprints, namely  Molecular ACCess System (MACCS) and Electrotopological state (Estate) fingerprints. The proposed model utilizes Pseudo amino acid composition derived using amino-acid sequences of targets via \emph{iFeature webserver}~\cite{chen2018ifeature}. We use drug Simplified Molecular-Input Line-Entry System (SMILES) to obtain CDK descriptors, MACCS fingerprints and Estate fingerprints. The purpose here is to evaluate the impact of employing CDK descriptors for DTI prediction. We compare models' performance against two frequently used fingerprints, MACCS fingerprints and Estate fingerprints on benchmark data. This work mainly focus on extracting and feature processing, followed by a systematic prediction methodology based on machine learning. For example, in this case, we utilize the Categorical Boosting (CatBoost) classifier to make predictions. For validation, we compare our proposed model to several recently proposed models. The results reveal that the proposed DTI prediction model identifies drug-target interactions more accurately using CDK descriptors than MACCS and Estate fingerprints. \\
We organize the paper as follows. Section~\ref{sec2} offers an overview of computation approaches to DTI predictions and highlights recent methods closely related to our work. Section~\ref{sec3} provides the details about the datasets and feature encodings. Section~\ref{sec4} describes our proposed methodology and a brief overview of the CatBoost algorithm. Evaluation metrics and performance results are presented in Section~\ref{sec5} and Section~\ref{sec6} respectively. Finally, we conclude the paper in Section~\ref{sec7}.
\section{Computation Approaches to DTI Predictions}\label{sec2}
In this section, we provide an overview of computation approaches to DTI predictions and highlight some closely related work to our proposed methodology. The computational strategies for the prediction of DTIs can be broadly divided into \textbf{ligand-based}, \textbf{docking-based}, and \textbf{chemogenomic} approaches~\cite{jacob2008protein}.

\textbf{Ligand-based. }The rationale behind ligand-based techniques is that identical compounds bind to identical biological targets and have identical features. It starts with a single molecule or a group of chemicals known to be effective against the target and it is further guided by the structure-activity relationships. However, there are certain drawbacks to this strategy. Because the protein sequence information is not employed in the prediction process, discovering new interactions reduces the connection across the identified ligand and protein families ~\cite{ezzat2019computational}. 

\textbf{Docking-based. }The docking-based approach, on the other hand, uses the 3D shape of proteins and chemical compounds to determine their possibilities of interaction ~\cite{li2006tarfisdock,cheng2007structure}.  However, specific proteins like the membrane proteins have unknown 3D structures that make it less applicable~\cite{opella2013structure}.
\begin{center}
\begin{figure}[t]
    \includegraphics[width=\textwidth]{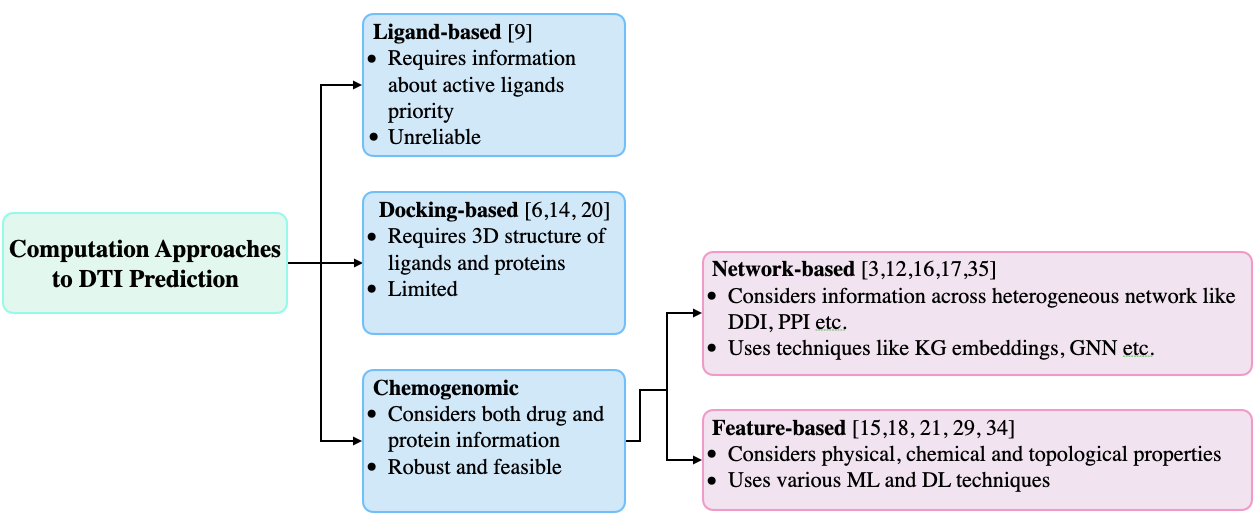}
    \caption{A brief taxonomy of computational approaches to DTI prediction}
    \label{Figure1}
\end{figure}
\end{center}
\vspace{-10mm}

\textbf{Chemogenomic. }The chemogenomic approaches use drug and protein information together to anticipate interactions. To infer probable interactions, a shared subspace is created by unifying the biochemical space of drugs and the genome space of targets. The main benefit of this method is that it utilizes a significant amount of biological data that is freely accessible from public repositories~\cite{yamanishi2008predictionf}. Chemogenomic approaches are roughly divided into \textbf{network-based} methods and \textbf{feature-based} methods. \textbf{Network-based} approaches integrate data like drug-drug interactions, protein-protein interactions, drug-disease interactions, and drug-target interactions from multiple sources into a single unified framework to boost DTI prediction ~\cite{chen2020prediction,ji2020prediction,liu2016neighborhood,luo2017network,zheng2013collaborative}. For instance, Wan et al.~\cite{wan2019neodti} devised an end-to-end technique entitled \emph{NeoDTI} to combine data from omics networks and learn topology that preserves the information of drugs and targets. Recent years have seen a fast growth of ML models based on knowledge graphs (KG). Mohammad et al.~\cite{mohameddiscovering} suggest \emph{triModel}, a model based on  Knowledge Graph (KG) embeddings to derive novel DTI from the model's scores built by learning embeddings about drugs and targets from multi-modal heterogeneous data. On the other hand, \textbf{feature-based} approaches represent each drug target pair as an array of descriptors. Drugs and proteins are transformed into corresponding descriptors based on their chemical properties. Integration of individual features of drugs and targets forms the input to these approaches as a 1D array ~\cite{li2021computational,pan2021prediction,yamanishi2010drug}. 

Most researchers prefer feature-dependent computation models to predict DTIs focussed on structural information of drugs based on molecular fingerprints that are bit strings indicating the existence of a specific substructure. For example, Han et al.~\cite{shi2019predicting} present an automatic learning system called \emph{LRF-DTIS} retrieving drug and target characteristics in the form of \emph{ PsePSSM} and molecular fingerprints employing lasso for feature selection, smote for handling imbalance, and random forest (RF) for predicting interactions. Wenyu et al.~\cite{mahmud2020prediction} suggest \emph{PDTI-ESSB} turning all drug molecules into molecular substructure fingerprints and representing protein sequences as multiple features to express their evolutionary, sequential, and structural information. To prevent the drawbacks of sparseness and dimensionality curse, Wang et.al.~\cite{wang2022mspedti} propose \emph{MSPEDTI} which uses a Convolutional Neural Network to derive relevant low-dimensional features from the sequence and structural information defined in the form of PSSM and molecular fingerprints. Likewise, Wang et al.~\cite{wang2022rofdt} propose a similar method using \emph{PsePSSM} and \emph{PubChem} fingerprints with feature weighted Rotation Forest. Sajadi et al.~\cite{sajadi2021autodti++} introduce an interesting approach to handling the sparsity in the interaction matrix of drugs and targets through drug fingerprints.
Wang et al.~\cite{wang2020predicting} use \emph{MACCS} and \emph{PAAC} for predicting DTIs with a novel method to create negative samples. Another way of capturing the structural information is molecular descriptors which are theoretically derived properties representing the physical, chemical, and topological characteristics of drugs. 
The majority of researchers use benchmark data to assess the effectiveness of their methods. We utilize the suggested benchmark data to evaluate our proposed method. 
In Section~\ref{sec3}, we give a brief description about the data and feature representation for drugs and proteins.

\section{Materials and Methods}\label{sec3}
\subsection{Dataset Description}
The gold-standard datasets used in this investigation are enzyme, Ion Channel (IC), Nuclear Receptor (NR) and G-Protein Coupled Receptor (GPCR), have been compiled by Yamanishi et al.~\cite{yamanishi2008predictionf} from the SuperTarget~\cite{gunther2007supertarget}, BRENDA~\cite{schomburg32amp}, DrugBank~\cite{wishart2008drugbank} and KEGG ~\cite{kanehisa2006genomics} repositories. The number of DTI pairs in these datasets after deleting the redundant information are $2926$, $635$, $1476$, and $90$, respectively. We consider each of these combinations as a positive interaction. To ensure a balance between positive and negative interactions, random under-sampling is applied to all pairs of DTIs to generate negative interactions. Table~\ref{tab1} displays the statistical data regarding these benchmark datasets.
\begin{table}[!t]
\caption{Qualitative details of the datasets.}\label{tab1}
\begin{center}   
\begin{tabular}{p{6pc}|p{3pc}|p{3pc}|p{7pc}|p{7pc}}
\toprule
\textbf{Dataset} & \textbf{Drugs} & \textbf{Targets} & \textbf{Known Interaction} & 
\textbf{Ratio of Imbalance}\\
\midrule
\textbf{Ion channel} &201 &204 & 1476 &0.036\\
\textbf{Enzyme} &445 &664 &2926 &0.010  \\
\textbf{Nuclear Receptor} &54 &26 &90 &0.068 \\
\textbf{GPCR} &223 &95 &635 &0.031 \\
\bottomrule
\end{tabular}
\end{center}
\end{table}
\subsection{Drug Feature Representation}
In this study, molecular descriptors and molecular fingerprints  convert the molecular structures of drugs into numerical form. For molecular descriptors, we use CDK descriptors derived from \addtocounter{footnote}{-2}ChemDes\footnote{\url{http://www.scbdd.com/chemdes/}}, for molecular fingerprints, namely MACCS and Estate, we use the RDKit library\footnote{\url{https://www.rdkit.org/docs/GettingStartedInPython.html}}. The concept behind fingerprints is to define molecular structure using a library of molecular substructures, that translates a chemical compound into a bit vector of 0's and 1s by detecting whether a compound has a  particular substructure or not. If the compound has that substructure, then only the bit at that position is put to 1 which results in a characterization of the molecular structure as a binary string. The number of molecular substructures in MACCS and Estate is 166 and 79 respectively, which is the size of the final binary bit vector. CDK descriptors, on the other hand, is an open-source platform for detecting and categorizing compounds using descriptor classes. The final size of the CDK descriptors is 275 and consists of autocorrelation descriptors, connectivity descriptors, constitutional descriptors, kappa descriptors, molecular properties, topological descriptors, WHIM descriptors, CPSA descriptors, geometrical descriptors, and quantum chemical descriptors. 

\subsection{Protein Feature Representation}
PseAAC is a parallel-correlation-based method for delineating protein information that generates 20 + D features. There are various models based on amino acid compositions (AAC) that lack target sequence-order knowledge. Pse-AAC, introduced in \cite{chou2001prediction} can be used to express both AAC and AA sequence order data. This technique is common in bioinformatics and related areas. Pse-AAC combines the core features of AAC with certain additional parts indicating a set of protein correlation factors and helps to improve model performance for multiple tasks. The expression for the features of PseAAC is as shown in Eq.~\ref{eq1}.
\\

\begin{equation}\label{eq1}
\centering
 P = [p_1, p_2, p_3, ... p_{19}, p_{20},..., p_{20+\lambda}],\quad (\lambda < S) \ ,
\end{equation}
\\
where S corresponds to the size of the target sequence under consideration. The components are represented as shown below:\\
\begin{equation}\label{eq2}
    P_j  =\frac{F_j}{\sum_{j=1}^{20} F_j+ W\sum_{k = 1}^{\lambda } \psi_j},  \quad 1<<20 \ ,\\
\end{equation}
\\
\begin{equation}\label{eq3}
    P_j  =\frac{W\psi_j}{\sum_{j=1}^{20} F_j+ W\sum_{k = 1}^{\lambda } \psi_j},\quad 20+1<<20+\lambda \ .
\end{equation}
\\
\noindent
Here, P stands for a vector of features, and W stands for a weight factor of 0.05. $F_j$ shows amino acid occurrence frequency, $\psi$ is the sequence correlation factor and $\lambda$ represents information about the protein sequence order. In Eq.~\ref{eq2}, for a single protein, AAC is expressed by 20 components, whereas sequence order is expressed by $20+1$ to $20+\lambda$ components, referred to as Pse-AAC shown in Eq.~\ref{eq3}. Finally, for each protein sequence, we obtained a 50-D feature vector by setting the value lambda to 30. Rcpi~\cite{cao2015rcpi}, iFeature~\cite{chen2018ifeature}, and iLearn~\cite{chen2020ilearn} are among the useful software tools for encoding protein sequences. Section 4 defines the work flow of DTI prediction and shows the use of PAAC in our proposed methodology.
\begin{center}
\begin{figure}[t]
    \includegraphics[width=12cm, height=7cm]{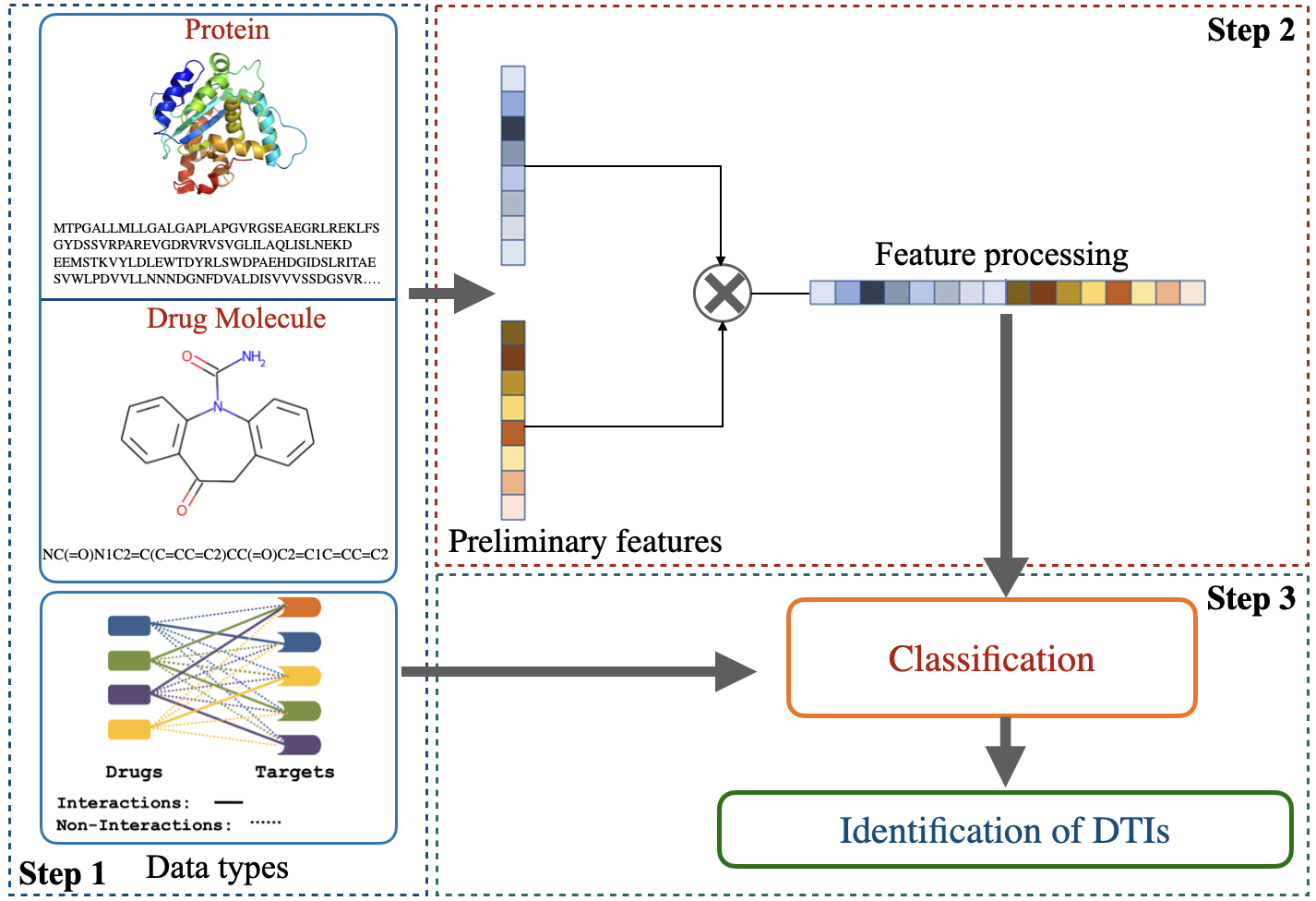}
    \caption{Proposed Methodology}
    \label{Figure2}
\end{figure}
\end{center}
\section{Proposed Methodology}\label{sec4}
The proposed workflow of our DTI prediction model is as shown in Fig.~\ref{Figure2}. It is comprised of three main steps. The description of each step is as follows:\\
\begin{itemize}
    \item \textbf{Step 1.} At first, we transform data into the corresponding values for which we represent each drug and target using their SMILES and amino acid sequences, respectively. Further, we use them to encode drug features and target features. For drugs, we use CDK descriptors, MACCS fingerprints and Estate fingerprints as features. For target protein, we use encoded features in the form of Pseudo amino acid composition (PseAAC).
    
    \item \textbf{Step 2.} In the next step, we process features from the input data. For each DTI pair, the drug features are concatenated with the target features to produce the final 1D array of features. The size of the array varies according to the drug feature under consideration. For instance, in the case of CDK descriptors, the length turns out to be $325$. For MACCS and Estate fingerprints, it is $216$ and $129$, respectively. To normalize the range of independent attributes, we use the StandardScaler() function that converts the training and testing data to scaled representation. It accelerates calculations in machine learning techniques.
    
    \item \textbf{Step 3.} In the final step, we provide the processed features and existing DTIs to a classification model as input. We use the CatBoost algorithm to train a classifier. After
    this stage, we measured the performance of the trained model on a test dataset. We provide further details about the CatBoost algorithm in the following subsection.
\end{itemize}
    
\subsection{CatBoost}
Gradient boosted trees and Random forest is robust machine models for tabular heterogeneous data.
CatBoost classifier is another open-source gradient boosting framework released in 2017~\cite{prokhorenkova2018catboost}. Although mainly designed to handle category features, CatBoost works on numerical and text data. According to the literature, it outperforms boosting algorithms such as XGBoost and LightGBM on a variety of datasets and has a substantially shorter prediction time.
The technique is well-known for its use of ordered boosting to counteract overfitting and the use of symmetric trees for faster execution. During model training, a sequence of decision trees is built one after the other, with each succeeding tree having a smaller loss. In other words, each decision tree learns from its predecessor and influences the subsequent trees to increase its performance, culminating in a powerful learner. Gradient boosting trees are good at dealing with numerical data but struggle with categorical features. Both strategies require a large amount of memory and are computationally expensive. CatBoost, as a solution, uses target-based statistics to address categorical features, saving time and resources. To overcome overfitting, the CatBoost method employs an ordered boosting mechanism. After numerous boosting stages, traditional gradient boosting approaches use all the training samples to develop a prediction model. This strategy causes a prediction shift in the created model, resulting in a distinct form of target leakage problem. CatBoost eliminates the aforementioned challenge by utilizing an ordered boosting architecture. Furthermore, unlike standard learning classifiers, the CatBoost approach gracefully handles overfitting by employing many permutations of the training dataset, which emerges as the main reason for deploying its intelligence in the current work. Five-fold cross-validation is used to avoid data overfitting and for the appropriate application of CatBoost in the current DTI problem. We evaluate the performance of the created model on multiple performance metrics defined in the next section.

\section{Evaluation Parameters}\label{sec5}
To evaluate and compare our proposed methodology with other methods, we use different metrics such as Precision, Sensitivity (or Recall), Accuracy, Mathews Correlation Coefficient (MCC), Area Under Curve (AUC), and Area Under Precision-Recall( AUPR). The AUC and AUPR graphs are good choices for unbalanced data. Hence, most research uses it as a comparison criterion. The ROC curve demonstrates how well the trained model performs at different cutoffs. False positive rates are compared with actual true positive rates to form the curve. The AUC values range from 0 to 1, with higher values suggesting an effective model. As the AUC summarizes the curve with a range of cutoff values as a single score, the AUPR also does the same. The difference is that it shows the precision (y-axis) and recall (x-axis) for various probability cutoffs. The following are the definitions of other evaluation parameters used in this study:

\begin{align}
  Precision &= \frac{TP}{TP + FP}\ ,\\
  Sensitivity &= \frac{TP}{TP + FP}\ ,\\
  Accuracy &= \frac{TP + TN}{TP + FP + TN + FN }\ ,\\ 
  MCC &= \frac{TP * TN - FP * FN}{(TP + FP)(TN + FN)(TN + FP)(TN + FN)}\ ,
\end{align}

where TP means correctly labeled positive observations, FP implies wrongly labeled negative observations, TN means correctly labeled negative observations, and FN implies wrongly labeled positive observations.

\section{Performance Evaluation}\label{sec6}
\begin{center}
\begin{figure}
    \centering     
    \begin{subfigure}[b]{0.45\textwidth}
        \centering
        \includegraphics[width=5.5cm, height=3.5cm]{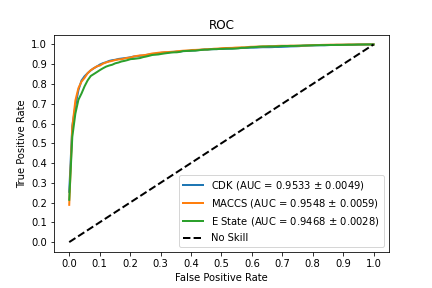}
        \caption{}
    \end{subfigure}
    \begin{subfigure}[b]{0.45\textwidth}
        \centering
        \includegraphics[width=5.5cm, height=3.5cm]{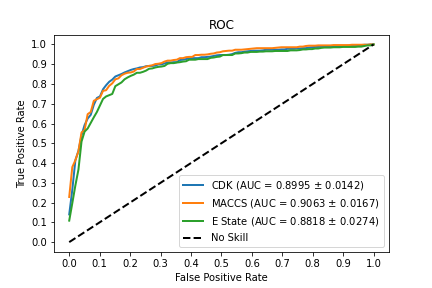}
        \caption{}
    \end{subfigure}
    \\
    \begin{subfigure}[b]{0.45\textwidth}
        \centering
        \includegraphics[width=5.5cm, height=3.5cm]{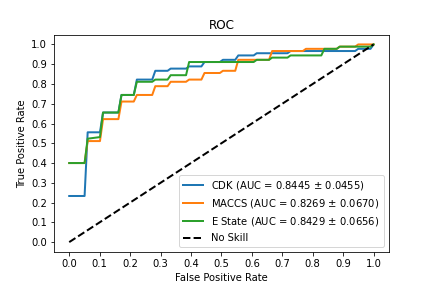}
        \caption{}
    \end{subfigure}
     \begin{subfigure}[b]{0.45\textwidth}
        \centering
        \includegraphics[width=5.5cm, height=3.5cm]{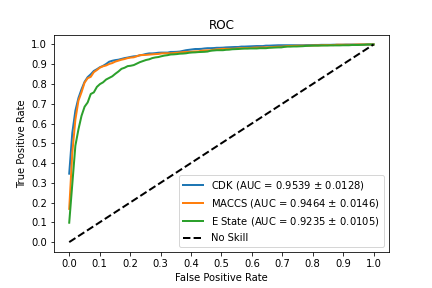}
        \caption{}
    \end{subfigure}
\caption{Comparative analysis of the area under the ROC curves obtained using CDK descriptors, MACCS Fingerprints and Estate fingerprints across all datasets.}
\label{Figure3}
\end{figure}
\end{center}

To evaluate the prediction capacity of our proposed approach, we use 5-fold cross-validation on all DTI networks, including IC, Enzymes, NR, and GPCR. We roughly divide the data into 5-folds and use four folds for the model training and the left-out fold for testing the model's performance. We repeated the process five times to get five different result sets and reported the mean values. Table~\ref{tab2} displays the experimental findings of five-fold cross-validation on various performance evaluators across all datasets.
\begin{table*}[]
\vspace{-5mm}

    \centering
  \caption{Obtained results using CDK descriptors, MACCS fingerprints, and Estate fingerprints across all datasets for different performance metrics.\\}   
\begin{tabular}{p{4pc}|p{3pc}|p{3pc}|p{3pc}|p{3pc}|p{3pc}|p{3pc}|p{3pc}}
\toprule
\textbf{Dataset} &\textbf{Feature} & \textbf{Accuracy} &\textbf{Precision} &\textbf{Recall} &\textbf{MCC} &\textbf{AUC} &\textbf{AUPR}\\
\midrule
\multirow{3}{*}{\textbf{Enzyme}}  &\textbf{CDK} &\textbf{0.8988} &\textbf{0.9015} &\textbf{0.8954} &\textbf{0.7977} & 0.9533 & \textbf{0.9597}\\\cline{2-8}
    & \textbf{MACCS} & 0.8975 &0.9013 &0.89262 &0.7915 &\textbf{0.9548} &0.9595\\\cline{2-8}
    & \textbf{Estate} & 0.8822 &0.8854 &0.8783 &0.7646 &0.9468 &0.9521\\\cline{2-8}
    \hline
\multirow{3}{*}{\textbf{GPCR}}& \textbf{CDK} & \textbf{0.8409} & \textbf{0.8304} & \textbf{0.8567} & \textbf {0.6826} & 0.8995 & 0.8967\\\cline{2-8} 
& \textbf{MACCS} & 0.8314 &0.8178 &0.8551 &0.6650 &\textbf{0.9063} &\textbf{0.9029}\\\cline{2-8}
& \textbf{Estate} & 0.8196 &0.8104 &0.8346 &0.6400 &0.8818 &0.8713
   \\\hline
\multirow{3}{*}{\textbf{IC}}& \textbf{CDK} & \textbf{0.8929} & \textbf{0.8864} & \textbf{0.8997} &\textbf{0.7848} &\textbf{ 0.9539} &\textbf{0.9591}\\\cline{2-8} 
& \textbf{MACCS} &0.8872 &0.8793 &0.8977 &0.7750 &0.9464  &0.94767\\\cline{2-8}
&\textbf{EState} & 0.8523 &0.8409 &0.8692 &0.7058 &0.9235 &0.9174
   \\\hline   
\multirow{3}{*}{\textbf{NR}}& \textbf{CDK} & \textbf{0.8056} &0.7788&\textbf{0.8556} & \textbf{0.6155} & \textbf{0.8445} & \textbf{0.8376}\\\cline{2-8} 
& \textbf{MACCS} & 0.75 &0.7560 &0.7778 &0.5101 &0.8269 &0.8342\\\cline{2-8}
& \textbf{EState} & 0.7611 &\textbf{0.7837} &0.7333 &0.5332 &0.8429 &0.8327
   \\\bottomrule  
\end{tabular}

    \label{tab2}
\end{table*}
Table~\ref{tab2} shows that we obtain the best accuracy of 0.8988 and the second best accuracy of 0.8929 using CDK descriptors on the enzyme and ion channel dataset, respectively. Results also show that the performance of CDK descriptors is better in most cases compared to MACCS and Estate fingerprints, which confirms that CDK descriptors are better at predicting the DTIs compared to these two widely used molecular fingerprints. Fig.~\ref{Figure3} shows the performance curves of the area under the ROC curve across all benchmark datasets.
\\
\vspace{-15mm}
\subsection{Comparison with previous methods}
To evaluate the proposed model's capacity for predicting DTIs in a logical manner, we contrasted it with earlier techniques using the gold-standard dataset and chose the AUC as the assessment measure since it best captures the model's performance. Table~\ref{tab3} aggregates the AUC values from prior approaches such as Yamanishi ~\cite{yamanishi2010drug}, DTCWT~\cite{pan2021prediction}, Yang~\cite{li2021computational}, Elastic net~\cite{shi2019predicting}, RoFDT~\cite{wang2022rofdt}, AutoDTI++~\cite{sajadi2021autodti++} and MSPEDTI~\cite{wang2022mspedti}. Table~\ref{tab3} clearly shows that the proposed model outperformed the prior technique across benchmark datasets. This shows that using CDK descriptors with the CatBoost classifier can significantly improve the capacity to anticipate DTIs.
\begin{table}
\caption{Comparison of the proposed methodology with previous studies with respect to AUC values.}\label{tab3}
\begin{center} 
\begin{tabular}{p{7pc}|p{5pc}|p{5pc}|p{5pc}|p{5pc}}
\toprule
\textbf{Methods}  & \textbf{Enzyme}  &\textbf{GPCR} &\textbf{IC} &\textbf{NR}\\
\midrule
\textbf{Yamanishi}~\cite{yamanishi2010drug} &0.821 &0.811 &0.692 &0.814\\
\textbf{Elastic net}~\cite{shi2019predicting} &0.8605 &0.7785 &0.804 &0.8418\\
\textbf{Yang}~\cite{li2021computational} &0.9529 &0.8878 &0.925 &\textbf{0.8487}\\
\textbf{DTCWT}~\cite{pan2021prediction} &0.9498 &0.8775 &0.9270 &0.7755\\
\textbf{RoFDT}~\cite{wang2022rofdt} &0.9172 &0.8557 &0.8827 & {0.7531}\\
\textbf{AutoDTI++}~\cite{sajadi2021autodti++} &0.90 &0.86 &0.91 & \textbf{0.87}\\
\textbf{MSPEDTI}~\cite{wang2022mspedti} &0.9437 &0.8802 &0.9088 & \textbf{0.8663}\\
\textbf{Proposed Model} &\textbf{0.9533} &\textbf{0.8995} &\textbf{0.9539} &0.8445\\
\bottomrule
\end{tabular}
\end{center}
\end{table}
\section{Conclusion}\label{sec7}
We presented a machine learning-based prediction model for DTIs. We use random under-sampling to deal with the imbalance of the datasets. To encode features such as CDK descriptors, MACCS fingerprints, and Estate fingerprints for drugs and PseAAC for targets, we use drug chemical structures and amino acid sequences. The objective of this research is to evaluate the impact of employing CDK descriptors for DTI prediction. We compare its performance against two frequently used fingerprints, namely MACCS fingerprints, and Estate fingerprints. The experimental findings reveal that CDK descriptors outperform the other two commonly used fingerprints. We use five-fold cross-validation criteria to get the results. The provided methodology is both practical and effective in forecasting DTIs, according to the comparative outcomes. We intend to expand our research in the future by taking into account novel target feature descriptors coupled with deep learning techniques.

%
%

\end{document}